# On the Acoustoelasticity of Backward Lamb Waves in Prestressed Plates

Zhongtao Hu[1*], Guo-Yang Li[2*], and Hanyin Cui[3]

[1] *School of Engineering Medicine, Beihang University, Beijing 100191, China*

[2] *Department of Mechanics and Engineering Science, College of Engineering, Peking University, Beijing 100871, China*

[3] *State Key Laboratory of Acoustics, Institute of Acoustics, Chinese Academy of Sciences, Beijing 100190, China*

**Abstract:**

Backward Lamb waves, which exhibit a group velocity that propagates in the opposite direction to their phase velocity, have recently garnered considerable attention for their potential applications in nondestructive testing. Herein we present a theoretical study on backward Lamb waves in the elastic plate subject to prestresses. We demonstrate that the group velocity of the first antisymmetric backward Lamb wave, $A_{3b}$, decreases with tensile stress, whereas that of the first symmetric backward Lamb wave, $S_{2b}$, increases. Notably, the sensitivity of $A_{3b}$ to prestress is approximately ten times greater than that of $S_{2b}$, with a ~5% change in group velocity observed under a uniaxial stress of 100 MPa in steel. This heightened sensitivity facilitates an inverse method for determining prestress levels in elastic plates by examining variations in the $A_{3b}$ group velocity. We also investigate the acoustoelastic properties of zero-group-velocity (ZGV) points, which demarcate the dispersion curves of forward and backward Lamb waves. Our findings indicate that the ratio of resonance frequencies corresponding to $A_{3b}$ and $S_{2b}$ monotonically decreases as uniaxial stress increases, providing an alternative method for prestress assessment. Lastly, we propose an experimental setup for measuring backward Lamb waves and visualize the generation of $A_{3b}$ using dynamic photoelastic techniques. Our research elucidates the acoustoelastic characteristics of backward Lamb waves and highlights their promising utility for stress measurement in elastic plates.

[*] Corresponding author:

Guo-Yang Li (lgy@pku.edu.cn), Zhongtao Hu(zhongtaohu@buaa.edu.cn)

# 1 Introduction

Backward Lamb waves, characterized by the unique property of phase and group velocity moving in opposite directions, have piqued considerable interest in diverse areas of wave physics [1–10]. This enigmatic behavior was first delineated by Tolstoy et al. [11], who elucidated that in isotropic plates, Lamb waves could manifest with their phase front and energy flux traveling in diametrically opposite directions. Experimental observations of backward waves were first noted in elastic cylinders and plates [12]. Subsequently, various experimental methods have been proposed to investigate backward Lamb waves [13,14]. Notably, the propagation of backward Lamb waves in plates can be intuitively visualized using dynamic photoelastic techniques [15–17]. Recent research has explored the applications of backward Lamb waves, including negative reflection at plate edges [18–20], achieving negative refraction via mode conversion in plates with step-like discontinuities [21,22], controlling beam diffraction effects in fluid-embedded plates under normal incident waves [23], and highly precise non-contact estimation of concrete plate thickness [24].

The frequently studied backward Lamb wave modes in experiments are $S_{2b}$ and $A_{3b}$, with 'S' and 'A' denoting symmetric and anti-symmetric Lamb wave modes, respectively, and 'b' representing backward wave propagation. In the real wavenumber domain, the dispersion relations of $S_{2b}$ and $A_{3b}$ are connected to those of $S_1$ and $A_2$ at points where zero-group velocities (ZGVs) emerge while phase velocities remain finite [25]. Historically, these ZGVs are referred to as $S_{1\text{-ZGV}}$ and $A_{2\text{-ZGV}}$. ZGV modes localize their energy, making them valuable for nondestructive evaluation (NDE) of various physical properties, including wall thickness [26,27], elastic constants [28–30], and fatigue damage [31,32].

While backward Lamb waves have undergone extensive study, their interaction with prestress remains unexplored. Prestress is expected to influence wave propagation through the acoustoelastic effect, which, in turn, forms the basis for nondestructive evaluation (NDE) of stress [33–36]. The investigation into the acoustoelasticity of Lamb waves traces its origins to early work by Ogden [37]. In 2012, Gandhi et al. conducted a comprehensive analysis of Lamb wave propagation in elastic plates under biaxial loading conditions [38]. In search of enhanced sensitivity to prestress, subsequent studies by Mohabuth et al. and Ning et al. delved into the acoustoelasticity of higher-order Lamb modes [39,40]. Pau et al. developed an analytical model for nonlinear Lamb waves in prestressed plates [41]. Recent advancements in elastography have reignited interest in acoustoelastic Lamb waves, where these waves have been employed for mechanical characterization of soft materials [42–44]. However, current investigations have predominantly focused on the acoustoelastic effects associated with forward Lamb waves, while the interaction of backward Lamb waves with prestressed plates remains unexplored.

Taking all this into account, our study focuses on the acoustoelasticity of backward Lamb waves, with a primary emphasis on their potential for NDE of stress in elastic plates. We conducted a thorough theoretical analysis of the acoustoelastic properties of backward Lamb waves, revisiting incremental dynamics. Building on this analysis, we proposed novel approaches for probing stresses within elastic plates. Additionally, we introduce an experimental setup designed for measuring backward Lamb waves and visually demonstrating the generation of A3b using dynamic photoelastic techniques. In conclusion, our findings suggest that backward Lamb waves, displaying exceptional sensitivity to prestress, offer a promising avenue for accurately estimating stress levels within elastic plates.

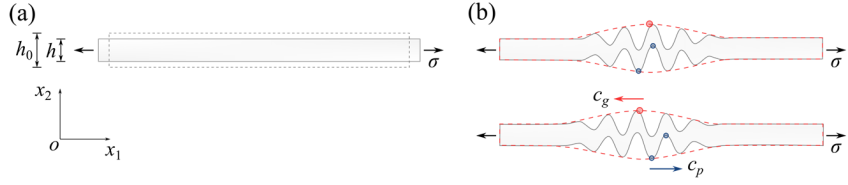

Fig. 1 **Schematic diagram of small-amplitude Lamb waves in a prestressed elastic plate**. (a) An elastic plate subject to a prestress field. The stress is assumed to be uniform locally. The principal stresses are $\sigma_1$ and $\sigma_2$. (b) The schematic of a backward Lamb wave within the cross-sectional view of the plate, in the principal coordinate system $x_1$–$x_2$. The phase velocity $c_p$ and group velocity $c_g$ are in opposite directions.

## 2 Acoustoelasticity of backward Lamb waves

The acoustoelasticity of Lamb waves have been widely studied and proved to be useful to probe mechanical properties and stresses in elastic membranes [37,42,45,46]. Here we simply revisit the dynamic incremental theory that can be utilized to derive the dispersion relations of acoustoelastic Lamb waves.

To begin with, we consider an elastic plate, with an initial thickness $h_0$, is subject to a uniaxial stress $\sigma_{11}$ along $x_1$ axis (see Fig. 1a). The prestress changes the thickness to $h = \lambda_2 h_0$, where $\lambda_2$ is the stretch ratio along $x_2$ axis. To relate the stress ($\sigma_{ii}$) to the deformation of the plate ($\lambda_i, i = 1,2,3$), we should introduce the strain-energy function $W$. Here we take the third-order strain energy function that is expressed in terms of the right Cauchy-Green strain tensor $\mathbf{C}$ [47]

$$W = \frac{\lambda}{8}(I_1 - 3)^2 + \frac{\mu}{4}(I_1^2 - 2I_1 - 2I_2 + 3) + \frac{l}{24}(I_1 - 3)^3 + \frac{m}{12}(I_1 - 3)(I_1^2 - 3I_2) \\ + \frac{n}{8}(I_1 - I_2 + I_3 - 1). \tag{1}$$

The right Cauchy-Green tensor is defined as $\mathbf{C} = \mathbf{F}^T\mathbf{F}$, where $\mathbf{F}$ is the deformation gradient tensor, i.e., $\mathbf{F} = \text{diag}(\lambda_1, \lambda_2, \lambda_2)$. $\lambda$ and $\mu$ are the Lamé constants of linear elasticity. $l, m$, and $n$ are the second-order constants of Murnaghan [48]. $I_1, I_2$ and $I_3$ are three independent invariants of $\mathbf{C}$ defined as

$$I_1 = \text{tr}(\mathbf{C}), I_2 = \frac{1}{2}[I_1^2 - \text{tr}(\mathbf{C}^2)], I_3 = \det(\mathbf{C}). \tag{2}$$

Then the Cauchy stress tensor $\boldsymbol{\sigma}$ is related to the deformation by

$$\boldsymbol{\sigma} = J^{-1}\mathbf{F}\frac{\partial W}{\partial \mathbf{F}}, \tag{3}$$

where $J = \det(\mathbf{F})$.

In the deformed plate, the wave equation that governs small-amplitude wave motions is [49]

$$A_{0piqj}u_{j,pq} = \rho u_{i,tt}, \tag{4}$$

where $u_i$ denotes the incremental displacement. $\rho = J^{-1}\rho_0$, where $\rho_0$ denotes the density in the undeformed configuration. $A_{0piqj}$ is the Eulerian elasticity tensor defined as

$$A_{0piqj} = J^{-1} F_{p\alpha} F_{q\beta} \frac{\partial^2 W}{\partial F_{i\alpha} \partial F_{j\beta}}. \tag{5}$$

The incremental stress introduced by the wave motion is

$$s_{0ij} = A_{0piqj} u_{j,q}. \tag{6}$$

The surface of the plate should be traction free, i.e.,

$$s_{022} = 0, s_{021} = 0. \tag{7}$$

For the Lamb waves in $x_1 - x_2$ plane, we can write $u_i = U_i(x_2)e^{i(kx_1-\omega t)}$, where $U_i$ is the wave amplitude ($U_3 = 0$), $k$ is the wavenumber, and $\omega = 2\pi f$ is the angular frequency. Inserting the displacement into Eqs. (4) and (7), we can get a system of linear equations. To make sure there is nontrivial solution to the equations, we can finally get a secular equation that determines the dispersion relations of the Lamb waves. Detailed dispersion equation can be found elsewhere, for example in Ref. [40].

## 3 Prestress shifts the frequencies of zero-group-velocity (ZGV) Lamb waves

To demonstrate the impact of prestress on the backward Lamb waves, we consider a 1 mm thick rail steel plate. The elastic and acoustoelastic parameters of the steel can be found in Table 1. Figure 2(a) depicts the dispersion relation of the Lamb waves, i.e., the frequency $f$ as a function of wavenumber $k$. Different branches in this figure relate to different Lamb wave modes. Essentially, Lamb wave modes fall into two categories: antisymmetric modes ($A_0$, $A_1$, …) that feature antisymmetric displacements through the wall thickness, and symmetric modes ($S_0$, $S_1$, …) characterized by symmetric displacements. As the phase velocity $c_p$ and group velocity $c_g$ are determined by $2\pi f/k$ and $2\pi df/dk$, respectively, a negative slope in a dispersion relation means a negative group velocity, or a backward Lamb wave. In Fig. 2(a), two branches with negative slopes are marked out as red and blue solid line, respectively. The two backward Lamb wave modes are referred to as $A_{3b}$ and $S_{2b}$, since they connect to $A_3$ and $S_2$ modes by loops, respectively, in the imagery $k$ space [21]. The point where a backward branch connect to a forward branch corresponds to a zero-group-velocity (ZGV) point, i.e., $c_g = 0$. The ZGV points correspond to the $A_{3b}$ and $S_{2b}$ are marked in Fig. 1(a) with small circles. Since the two ZGV points to $A_2$ and $S_1$ as well, their frequencies are denoted by $f_{ZGV}^{A_2}$ and $f_{ZGV}^{S_1}$. It should be noted that the existence of the backward Lamb waves is primarily determined by the Poisson ratio $\nu = \lambda/(2\lambda + 2\mu)$. While the $S_{2b}$ exists in compressible materials ($\nu < 0.5$), $A_{3b}$ only occurs when $\nu < 0.32$ [28]. For rail steel (see Table 1), we find $\nu \approx 0.2959$.

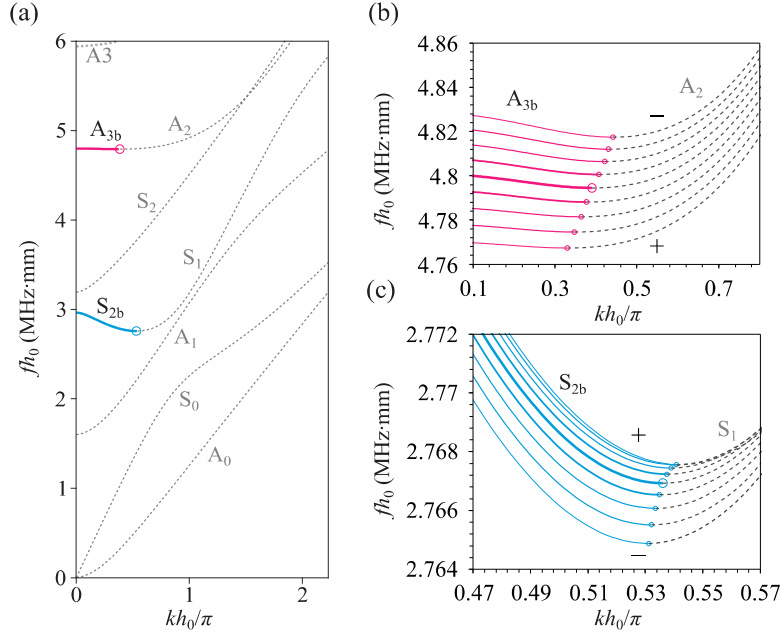

Fig. 2. **Effect of uniaxial stress on the backward Lamb waves.** (a) The dispersion relations of a rail steel plate (thickness $h_0 = 1$ mm). The $A_{3b}$ and $S_{2b}$ branches are marked out as red and blue solid lines, respectively. The effects of prestress on the (b) $A_{3b}$ and (c) $S_{2b}$. A uniaxial stress (−1 GPa to 1 GPa with step of 0.25 GPa) parallel to the wave propagation direction is applied. The zero-group-velocity (ZGV) points that correspond to $S_1$ and $A_2$ modes are marked as circles.

Then we consider the effect of prestress. Figures 2(b) and (c) present the dispersion relations of the $A_{3b}$ and $S_{2b}$, when the plate is subject to uniaxial stresses $\sigma_{11}$ from −1 to 1 GPa with a step of 0.25 GPa. We find the uniaxial stress shifts the $S_{2b}$ to higher frequency, whereas the $A_{3b}$ is shifted to an opposite direction. The shifts of the dispersion relations suggest the changes in ZGV frequencies. In Fig. 3(a), we plot the variations of $f_{ZGV}^{A_2}$ and $f_{ZGV}^{S_1}$. For a given material that is free from prestress, the product of the frequency and thickness for a ZGV mode should be constant. Therefore, the ratio changes in $h_0$ will result in the same ratio changes in the frequency. Due to the Poisson effect, a tensile/compressive stress will decrease/increase the thickness of the plate. Taking $f_{ZGV}^{S_1}$ as an example, we firstly investigate whether the geometry change introduced by the prestress dominates the variations in $f_{ZGV}^{S_1}$. By solving Eq. (3) we get the thickness varies linearly with the stress (±0.14% when $\sigma_{11} = \mp 1$ GPa). The variations in $f_{ZGV}^{S_1}$ that is purely introduced by the thickness change is shown in Fig. 3(a) (the dashed line). We find the stress significantly quenches the sensitivity of $f_{ZGV}^{S_1}$ to the thickness changes. From the thickness changes, we expect ±0.14% changes in $f_{ZGV}^{S_1}$. However, we get only 0.02% and −0.07% for $\sigma_{11} = 1$ GPa and −1 GPa, respectively. This result manifests the pronounced role of the acoustoelastic effect in determining the frequency of ZGV. For $f_{ZGV}^{A_2}$ we find a negative correlation with the stress, which is opposite to the contribution of the thickness change. This result further evidences the dominant role of the acoustoelastic effect and suggests the $A_2$ mode is more sensitive to the prestress than $S_1$ mode.

**Table 1.** Material parameters for four kinds of l steels. The density $\rho_0$ is in kg/m$^3$. The elastic and acoustoelastic constants are in GPa.

| Material | $\rho_0$ | $\lambda$ | $\mu$ | $l$ | $m$ | $n$ | $\alpha$ (GPa$^{-1}$) | $\beta$ |
|---|---|---|---|---|---|---|---|---|
| Rail steel [53] | 7800 | 115.8 | 79.9 | −248 | −623 | −714 | −9.901×10$^{-3}$ | 1.733 |
| Nickel-steel S/NVT [54] | 7800 | 109 | 81.7 | −56 | −671 | −785 | −12.51×10$^{-3}$ | 1.752 |
| Hecla ATV austenitic [55] | 7800 | 87 | 71.6 | −535 | −752 | −400 | −11.19×10$^{-3}$ | 1.774 |
| Hecla 37 (0.4%C) [55] | 7823 | 111 | 82.1 | −474.5 | −636 | −708 | −7.648×10$^{-3}$ | 1.749 |

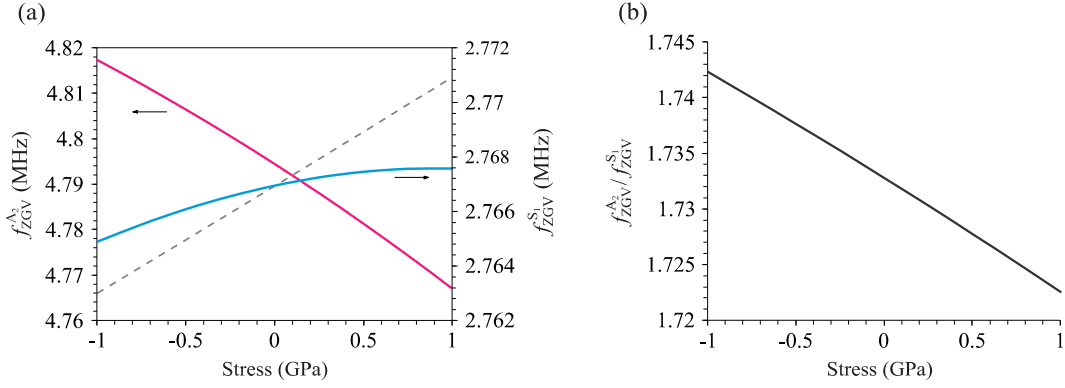

Fig.3. **Effect of uniaxial stress on the frequencies of the ZGV modes**. (**a**) Plots of the frequencies of the ZGV modes, $f_{ZGV}^{A_2}$ (Pink color) and $f_{ZGV}^{S_1}$ (Azure color), as functions of the uniaxial stress. The dashed line denotes the variations in $f_{ZGV}^{S_1}$ induced by thickness change. (**b**) Plot of the ratio $f_{ZGV}^{A_2}/f_{ZGV}^{S_1}$ as a function of the uniaxial stress.

For stress-free plate, the ratio $f_{ZGV}^{A_2}/f_{ZGV}^{S_1}$ is a function of the Poisson ratio, making it a candidate to precisely measure the Poisson ratio [28]. As shown in Fig. 3(b), we find the ratio $f_{ZGV}^{A_2}/f_{ZGV}^{S_1}$ monotonically decreases with the stress. We perform a linear fit to $f_{ZGV}^{A_2}/f_{ZGV}^{S_1}$ and find $f_{ZGV}^{A_2}/f_{ZGV}^{S_1} \approx \alpha\sigma_{11} + \beta$ ($r^2 > 0.999$), where $\alpha = -9.901 \times 10^{-3}$, $\beta = 1.733$, and $\sigma_{11}$ is in GPa. For all the steels listed in Table 1, we find the linear relationship holds and report the fitting parameters $\alpha$ and $\beta$.

## 4 Effect of prestress on the backward Lamb waves

We proceed to study the effect of prestress on the phase and group velocities of the backward Lamb waves. To this end, we compute the phase and group velocities from the dispersion relations shown in Fig. 2(a). As shown in Fig. 4(a) and (b), the group velocities of the backward Lamb waves are much lower than the phase velocities. In particular, the phase velocity of $A_{3b}$ is about 1000 times greater than its group velocity.

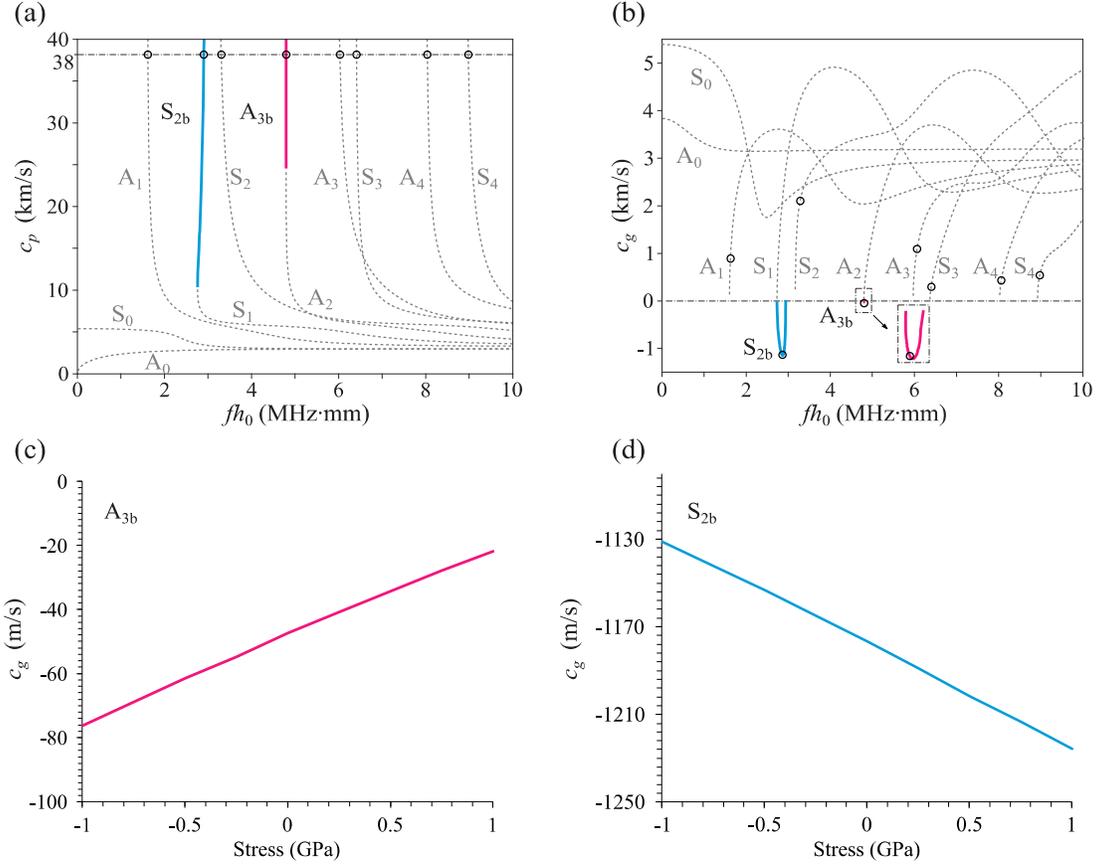

Fig. 4. **Effect of uniaxial stress on the phase and group velocities of the backward Lamb waves**. Plots of (a) phase and (b) group velocity as functions of the frequency. The dash-dotted line in (a) denotes a constant phase velocity 38 km/s. The intersections between the dash-dotted line and different Lamb wave modes are marked with circles. The group velocities corresponding to the intersections are marked in (b) with circles as well. (b) The group velocity curve of $A_{3b}$ is zoomed in for visualization. Variations in the group velocities of the (c) $A_{3b}$ and (d) $S_{2b}$ induced by a uniaxial stress.

Backward Lamb wave transports wave energy in the direction opposite to the forward Lamb wave. In addition, comparing to forward propagation branches, the backward wave branches are much fewer, and only existing within a narrow frequency range. The properties of backward Lamb wave make them advantageous for stress measurements in practical applications. For instance, the wedge-contact technique can generate multiple Lamb wave modes using pulsed ultrasound planar waves. The incident angle for these waves is set by an oblique wedge and is determined through Snell's law based on the chosen phase velocity and material properties [50]. Because of their opposing group velocities, backward Lamb waves can be isolated from forward Lamb waves, allowing for the selective capture of pure backward waves. When a constant phase velocity of 38 km/s is set using the wedge-contact technique (indicated by the red dashed line in Fig. 4(a)), specific Lamb wave modes are generated at intersections with this line, marked as red dots. The distribution of these intersections on group velocity dispersion curves is displayed in Fig. 4(b). Here, backward modes with negative group velocities propagate in the opposite direction to forward modes with positive velocities. Using a flat transducer with broad bandwidth and the wedge-contact technique, only two backward modes ($S_{2b}$ and $A_{3b}$) are acquired, with S2b propagating much faster.

Here, we quantitively analyze the variation of group velocity of backward Lamb waves with different uniaxial stress under the generated phase velocity of 38 km/s. The relative change in group velocity defined as $\Delta V/V_0$, where $\Delta V$ is the variation of group velocity under stress and $V_0$ is the group velocity under free stress, with the applied stress from -1GPa to 1GPa was calculated. The data for the cases of $A_{3b}$ and $S_{2b}$ modes are presented in Fig. 5. The relative change of the group velocity for the $A_{3b}$ mode is -0.5122 GPa$^{-1}$, and that for $S_{2b}$ is 0.0467 GPa$^{-1}$. It can be found that the relative change of the group velocity for the $A_{3b}$ mode is much larger than $S_{2b}$ mode, and they react opposite to each other, under the same ranges of the applied stress. It means the group velocity for the $A_{3b}$ mode will change around 50% percent with applied stress of 1GPa, which is quite sensitive to the applied stress.

Lastly, a comparative analysis is performed on the relative changes in group velocity for four types of steel materials listed in Table 1 (see Figs. 5(a) and 5(b)). Overall, the results keep the consistence that the relative change of the group velocity for the $A_{3b}$ mode is much larger than the $S_{2b}$ mode, and they react opposite to each other, under the same ranges of the applied stress.

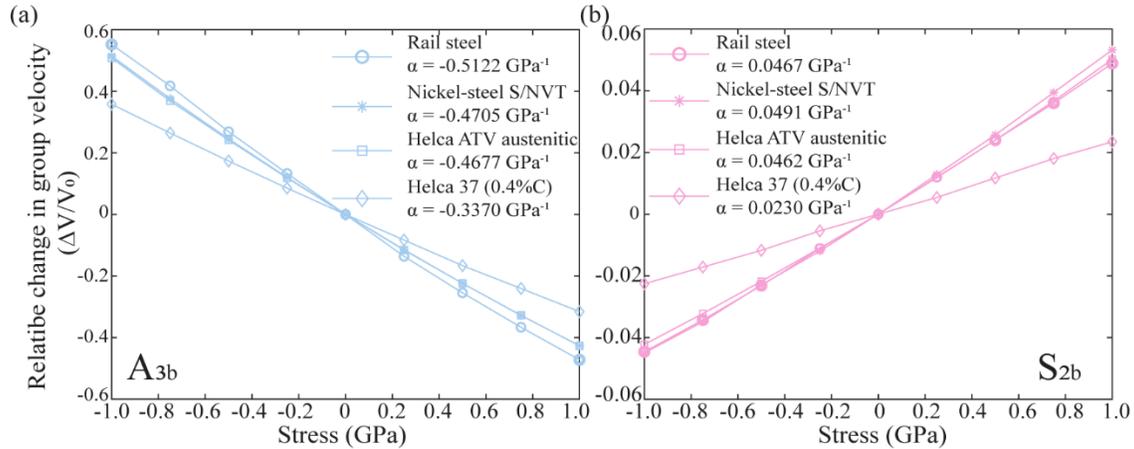

Fig. 5. The change of group velocity of backward Lamb wave with stress under constant phase velocity. (a) A3b and (b) S2b. The material constants of the four types of steels are listed in Table 1.

## 5 Experimental setup to measure backward Lamb waves

Dynamic photoelastic technique was used to verify the experimental method that generate backward Lamb wave in a glass plate. Figure 6(a) illustrates the dynamic photoelastic system designed for visualizing backward Lamb waves within a glass plate. Light emitted from a strobe YAG laser was expanded by a concave lens and then collimated by a collimating lens. The flashing duration of the laser source was 10 ns. This collimated light was incident normally onto the glass plate. Two linear polarizers (a polarizer and an analyzer) were positioned in a crossed arrangement. The glass sample is optically isotropic when unstressed and becomes anisotropic under stress due to a temporary birefringent effect. When the glass sample exhibits birefringence due to the stresses introduced by the Lamb wave, only the extraordinary ray produced by the birefringence can pass through the analyzer. This ray then transmits through a converging lens and an imaging lens before being captured by a CMOS camera and sent to a computer. A synthesizer controlled the time delay between the laser and the ultrasound transducer. By triggering the laser with a time delay relative to the ultrasound transducer, it became possible to observe and record the dynamic process of Lamb wave propagation inside the glass sample by adjusting the time delays.

Figure 6(b) showcases the setup for the generation of Lamb wave in the glass plate using the wedge-contact technique. The home-made ultrasound transducer has a diameter of 3 cm and a central frequency of 2 MHz with a 6 dB bandwidth of 1.4 MHz, allowing it to cover the frequency range from 1.3 MHz to 2.7 MHz. The plane wave emitted from the ultrasound transducer propagated along the 3 mm thick glass plate with the assistance of the polystyrene wedge, at a particular incident angle. The incident angle, $\theta$, is determined by the Snell's law and is calculated as $\theta = \arcsin(V_{ps}/C_{p0})$, where $V_{ps}$ is longitudinal wave speed of polystyrene wedge(2350 m/s) and $C_{p0}$ is the chosen phase velocity of Lamb waves. The material parameters of the polystyrene wedge are listed in Table 2.

The material parameters of K9 glass are detailed in Table 2. Figures 7(a) and (b) present the phase and group velocity dispersion curves for the K9 plate under stress-free conditions, respectively. The set phase velocity of the generated wave is 34,320 m/s, leading to an angle of 3.93° for the polystyrene wedge. The frequency-thickness product ($fd$) for the generated Lamb wave under the pulsed ultrasound wave is across from 3.9 to 8.1 MHz*mm. Points of intersection between the constant phase velocity and Lamb wave modes are marked as black hollow circles in Figures 7(a) and (b). Of these, four Lamb wave modes ($S_2$, $A_{3b}$, $S_3$ and $A_3$) are produced under pulsed ultrasound, and only $A_{3b}$ is the backward Lamb wave mode. A one-cycle pulsed ultrasound signal, centered at a frequency of 2 MHz, was generated by a Tektronix AFG 3102 signal generator. This signal was amplified by a high-power gated amplifier and then fed into the ultrasound transducer. Figure 8(a) shows the propagation of these generated Lamb waves inside the glass plate, as visualized by the dynamic photoelastic system. As time delays increase, wave energy spreads within the plate. A relatively stable wave, identified as backward Lamb wave $A_{3b}$, propagates in the negative direction. Conversely, some energy is dispersed in random directions, propagating forward due to the mixture of multiple forward Lamb modes. The propagation directions of the backward Lamb wave modes at each time step are indicated by black arrows. The numerical group velocity of $A_{3b}$ under the excited phase velocity of 34320 m/s is 514.8 m/s, which is consistence with the measured sound speed of 535.6 m/s ± 12.5 m/s calculated based on time-delays and propagation distance in photoelastic images. The results affirm that pulsed ultrasound, when propagated through a plate via an oblique wedge, can generate a pure backward Lamb wave mode.

We further substantiated our experimental findings through finite element (FE) simulations using Abaqus standard (Abaqus 6.13, Dassault Systèmes). To model the third-order elastic behavior, we developed a UHYPER user subroutine (Ref to [51]). The simulation domain consisted of a thin plate with a wall thickness of 3 mm and a wedge. The wedge was brought into contact with the prestressed plate, and we employed the same pulsed ultrasound stimulation as in the experiment to generate elastic waves in the wedge. These waves were then transmitted into the plate to induce selected Lamb waves. We ensured that the time increment was set to acquire more than 8 points in one wave cycle, and the size of the second-order, 4-node elements used in the analysis was smaller than 1/10 of the wavelength. Figure 8(b) presents the simulation results, clearly depicting the presence of a backward Lamb wave, with the vertical distribution of the map (magnitude of the particle velocity, v) indicating the predominant excitation of the $A_{3b}$ mode. This mode also appears in the negative direction, consistent with our experimental findings. The propagation distance in both the simulation (-27.3 mm) and the experiment (-26.7 mm) occurs within the same 50 μs timeframe. The wave energy distribution closely aligns with experimental data, thereby confirming the robustness of our experimental setup for generating A3b mode Lamb waves.

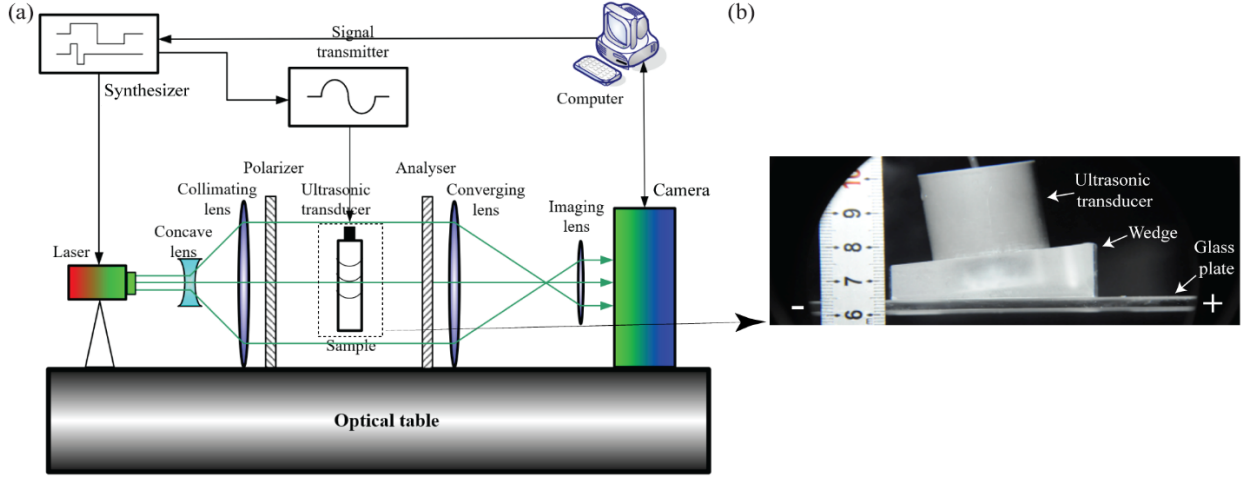

Fig. 6 (a) Dynamic photoelastic system for visualization of Lamb wave, and (b) the setup of Lamb wave generation in a glass plate.

| Material | $\rho_0$ | $\lambda$ | $\mu$ |
|---|---|---|---|
| K9 Glass[52] | 2510 | 24.2 | 33.1 |
| Polystyrene[17] | 1050 | 3.2 | 1.3 |

**Table 2.** Material parameters for k9 glass. The density $\rho_0$ is in kg/m$^3$ and the elastic constants are in GPa.

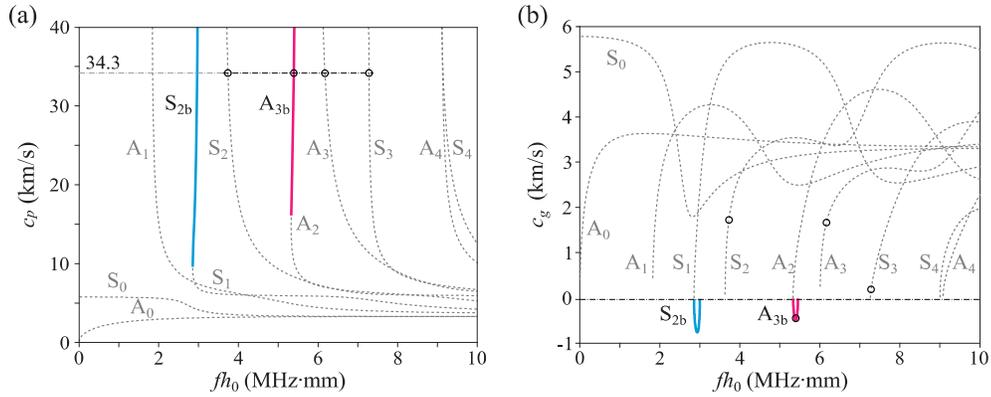

Fig. 7 Phase (a) and group (b) dispersion curved of k9 glass plate, and frequency-thickness product of the bandwidth of the ultrasound transducer.

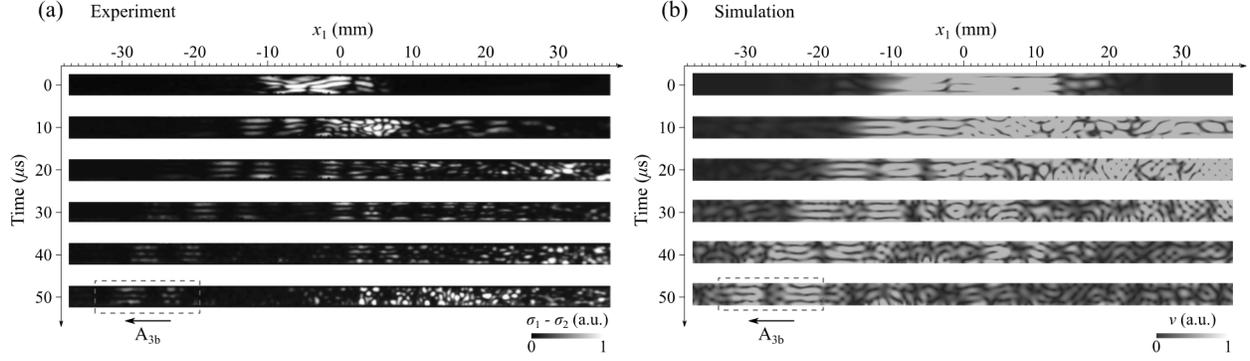

Fig. 8 (a) Backward Lamb wave generated in plate visualized using dynamic photoelastic system. (b) Finite element simulation of the experiment.

## 6  Discussions

In this study, we have made progress in clarifying the effects of acoustoelasticity on backward Lamb waves in prestressed plates. Our investigation sheds light on the behavior of the backward Lamb wave modes ($A_{3b}$ and $S_{2b}$) and their corresponding ZGV points under varying uniaxial stress conditions, uncovering distinctive anisotropic patterns and dispersive characteristics with valuable practical applications.

The study revisits and substantiates the relevance of dynamic incremental theory as a cornerstone for deriving the dispersion relations of acoustoelastic Lamb waves. We conclusively demonstrate that uniaxial stress differentially impacts $S_{2b}$ and $A_{3b}$ modes: $S_{2b}$ shifts toward higher frequencies, while $A_{3b}$ migrates in the opposite direction. More strikingly, based on the peculiar property of ZGV modes, we propose a dimensionless parameter $\bar{f}$ defined as $\bar{f} = f_{A2-ZGV}/f_{S1-ZGV}$, where $f_{A2-ZGV}$ and $f_{S1-ZGV}$ are the frequency of ZGV modes $A_{2\text{-}ZGV}$ and $S_{1\text{-}ZGV}$, respectively. Intriguingly, the parameter $\bar{f}$ reveals a monotonically decreasing trend with escalating levels of applied stress, thereby serving as a potential diagnostic metric for stress-state assessment.

From a practical perspective, the study also brought forth robust experimental and computational validations that attested to the efficacy of backward Lamb waves in stress measurement applications. Notably, the $A_{3b}$ mode displayed significantly higher sensitivity to uniaxial prestress compared to the $S_{2b}$ mode. A key innovation was the utilization of a wedge contact technique to selectively isolate the $A_{3b}$ mode, subsequently validated using dynamic photoelastic methods and numerical simulations. This experimental apparatus substantiates the potential of backward Lamb waves as a reliable tool for assessing stress states in isotropic materials.

However, a limitation of this study is its focus on isotropic and homogeneous materials. Further research is essential to explore the behavior of backward Lamb waves in more complex settings, such as composite and anisotropic materials. Future endeavors should delve into these uncharted territories to broaden the technique's applicability and reliability.

# 7 Conclusion

To conclude, this investigation represents progress in understanding the acoustoelastic effects on backward Lamb waves. It offers valuable theoretical insights and practical methodologies for applications in non-destructive evaluation and structural health monitoring. The robustness of our findings suggests opportunities for further research, particularly in the context of composite materials, and hints at the potential of backward Lamb waves in stress measurement applications. This study contributes to ongoing discussions and lays the groundwork for future investigations, which could lead to new perspectives on utilizing Lamb waves in engineering applications.

# 8 Funding

Guo-Yang Li acknowledges the financial support from the Fundamental Research Funds for the Central Universities, Peking University.